\documentclass[11pt,twoside]{article}
\usepackage{asp2010}

\resetcounters

\begin{document}

\allowtitlefootnote

\title{Superfluidity and entrainment in neutron-star crusts}
\author{N. Chamel$^1$, J.M. Pearson$^2$ and S. Goriely$^1$
\affil{$^1$Institute of Astronomy and Astrophysics, Universit\'e Libre de Bruxelles, CP 226, Boulevard du Triomphe, B-1050 Brussels, Belgium\\
$^2$D\'ept. de Physique, Universit\'e de Montr\'eal, Montr\'eal (Qu\'ebec), H3C 3J7 Canada}}


\begin{abstract}
Despite the absence of viscous drag, the neutron superfluid permeating the inner crust of a neutron star can still be
strongly coupled to nuclei due to non-dissipative entrainment effects. Neutron superfluidity and entrainment have been 
systematically studied in all regions of the inner crust of a cold non-accreting neutron star in the framework of the 
band theory of solids. It is shown that in the intermediate layers of the inner crust a large fraction of ``free'' neutrons 
are actually entrained by the crust. The results suggest that a revision of the interpretation of many observable astrophysical 
phenomena might be necessary. 
\end{abstract}

\section{Introduction}
\label{intro}

No sooner did Bardeen, Cooper and Schrieffer (BCS) publish their theory of superconductivity than Migdal speculated 
about nuclear superfluidity in the interior of neutron stars. This possibility was further investigated by Ginzburg 
and Kirzhnits in 1964. The discovery of the first pulsars and the observation of frequency glitches followed by 
very long relaxation times of the order of months provided very strong evidence for the presence of superfluids in
neutron stars. In particular, pulsar glitches are generally believed to be related to the neutron superfluid that 
permeates the inner crust of neutron stars at low enough temperatures~\citep[see][for a review]{lrr}. The core of a 
neutron star may also contain different kinds of superfluids and superconductors but their nature and their properties 
still remain very poorly known. The existence of a neutron superfluid in neutron-star crusts has recently found
support from the observation of the initial cooling in persistent Soft X-ray Transients~\citep{shternin07,brown09}. 
Superfluidity is also expected to play a role in other astrophysical phenomena like the quasiperiodic oscillations 
detected in the giant flares from Soft Gamma Repeaters~\citep{andersson09}.

\section{Nuclear superfluidity}

Most studies of nuclear superfluidity have focused on homogeneous neutron matter. As the temperature $T$ falls below some 
critical temperature $T_c$, neutrons with anti-aligned spins and zero orbital angular momentum form Cooper pairs like electrons 
in conventional superconductors. Because these pairs are bosons, they behave coherently on a very large scale. As a consequence, 
the neutron condensate can flow without any viscosity, analogous to superfluid helium-3. 

Recent microscopic calculations using realistic nucleon-nucleon interactions and based on different many-body methods tend to 
predict the same density dependence of the neutron pairing gap, at least at not too high densities~\citep{chang04,gezerlis2010}. 
In fact, the analytical expression of the gap in the limit of very dilute neutron matter has been known for a very long time~\citep{gorkov61}. 
These calculations predict a \emph{maximum} pairing gap $\Delta \sim 2$~MeV corresponding to a critical temperature  
$T_c=\exp(\gamma)\Delta/(k_{\rm B} \pi) \sim 10^{10}$~K where $\gamma$ is the Euler-Mascheroni constant and $k_{\rm B}$ the Bolzmann 
constant. 
 
However in spite of their names, neutron stars are not only made of neutrons. In particular, the neutron superfluid present in the 
inner crust coexists with neutron-rich clusters arranged in a regular Coulomb lattice and with a highly degenerate relativistic electron 
gas. Because the superfluid coherence length is much larger than the cluster size, the pairing phenomenon is highly non-local. This means 
that even if the center of mass of a neutron pair is located outside clusters, one of the partners may actually lie inside so that both 
bound and unbound neutrons are actually involved in the pairing phenomenon. As a result of these proximity effects, the properties of the 
neutron superfluid in the crust of a neutron star may differ substantially from those calculated in pure neutron matter. 

\begin{figure}
\centering
\includegraphics[scale=0.33]{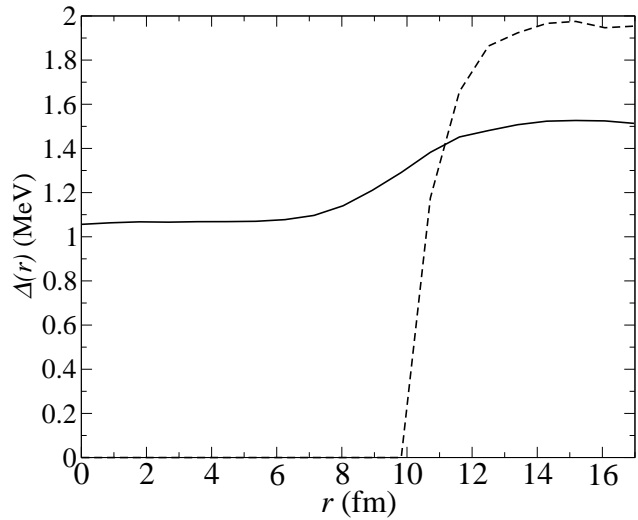}
\caption{Neutron pairing field in the Wigner-Seitz cell of the inner crust of a neutron star at the average baryon density 
$\bar n = 0.06$ fm$^{-3}$ and for $T=0.8 T_c$. Solid line: band structure calculation. Dashed line: at each point in the cell, the pairing field is approximated 
by its value in pure neutron matter for the corresponding local neutron density. See \cite{chapt2012} for further details. } 
\label{fig1}
\end{figure}

We have recently 
studied the inhomogeneous matter of the crust using the nuclear energy density functional (EDF) theory~\citep{cgpo2010}. This theory, which 
has been already very successfully applied to the description of medium-mass and heavy nuclei, provides a consistent description of both the 
neutron superfluid and the nuclear crystal lattice~\citep{chapt2012}. For this purpose, we have made use of accurately calibrated nuclear 
EDFs based on Skyrme effective nucleon-nucleon interactions complemented with a a microscopic pairing EDF~\citep{chamel2010}. The 
Brussels-Montreal EDFs~\citep{goriely2010} yield an excellent fit to essentially all experimental atomic mass data (2149 nuclei) with rms 
deviations lying below $0.6$~MeV, while reproducing at the same time i) a realistic neutron-matter equation of state and ii) the $^1S_0$ 
pairing gaps both in symmetric nuclear matter and in pure neutron matter, as obtained from many-body calculations including medium 
polarization effects. For these reasons, the Brussels-Montreal EDFs are well-suited for studying the pairing phenomenon in the inner 
crust of neutron stars. The neutron-star crust calculations have been carried out in the framework of the band theory of solids. In this 
way, the effects of the periodic crystal lattice on the neutron superfluid are properly taken into account. Because the pairing mechanism
directly arises from the strong attractive nuclear force, many neutrons are involved and not only those lying on the Fermi surface 
as in conventional superconductors. Neutrons belonging to different bands feel different pairing interactions thus leading to a dispersion 
of the pairing gap of a few hundred keV around the Fermi level. Moreover, pairing correlations are found to be enhanced in clusters while they 
are reduced in the intersticial region. These proximity effects become very important as the temperature approaches $T_c$, as illustrated 
in Fig.~\ref{fig1}. All in all, the critical temperature is found to be reduced by $\sim 10-20\%$ as compared to pure neutron 
matter~\citep{cgpo2010}. The effects of the crystal lattice on the superfluid dynamics are much more dramatic.

\section{Entrainment}

Even though the neutron superfluid can flow without friction, it can still be entrained by the crust~\citep{carter06,pethick2010}. 
Unlike viscous drag, this entrainment effect is non-dissipative. It arises because neutrons can be elastically scattered by the 
nuclear lattice. Since the first experiments by \cite{hal36} and ~\cite{mit36} a few years after the discovery of the neutron, the neutron 
diffraction has been routinely applied to study crystal structure. This is because neutrons reflected by parallel lattice planes spaced 
$d$ appart can interfere constructively but only for specific incident angles $\theta$ (measured from the plane) given by Bragg's law 
$2 d\sin\theta=N \lambda$, where $N$ is any integer and $\lambda\leq 2 d$ is the de Broglie wavelength of neutrons. Similarly, 
neutron diffraction can occur in the inner crust of a neutron star. Note that Bragg diffraction is a highly non-local phenomenon. 
Whereas neutrons used for diffraction experiments are produced in nuclear reactors and accelerators, the neutrons in the crust of a neutron 
star originate from electron captures on nuclei during the gravitational core-collapse of the supernova. With increasing matter density, 
nuclei thus become more and more neutron rich. As the matter density reaches about $4\times 10^{11}$ g~cm$^{-3}$, some neutrons drip out of 
nuclei and form a highly degenerate quantum liquid. A ``free'' neutron that is Bragg reflected cannot propagate and is therefore trapped in 
the crust: its velocity in the crust frame is zero. In other words, this neutron is entrained by nuclei. Even if a neutron is not Bragg 
reflected, its velocity will generally differ from that in empty space and will be approximately given by $\textbf{v}=\textbf{p}/m_n^\star$, where $\textbf{p}$ is the neutron momentum and 
$m_n^\star$ (which depends on $\textbf{p}$) is a neutron effective mass caused by the quantum mechanical interaction of the neutrons with 
the periodic lattice. For the momenta satisfying Bragg's law, the effective mass becomes infinitely large to the effect that $\textbf{v}=0$. 
Due to the Pauli exclusion principle, free neutrons have different momenta $\textbf{p}$ and are therefore diffracted differently. 
We have systematically calculated the momentum-averaged neutron effective mass in all regions of the inner crust using the band theory 
of solids~\citep{chamel2012}. The largest effects of Bragg diffraction are found in the intermediate parts of the inner crust at baryon 
densities $\bar n\sim 0.02-0.03$~fm$^{-3}$ where $m_n^\star\sim 14 m_n$. These results imply that only a very small fraction of neutrons are not
entrained by clusters and can therefore be considered as effectively ``free''. Alternatively, these results imply that clusters are effectively much heavier than generally 
assumed~\citep{pethick2010}. 

\section{Conclusions}

The pairing mechanism giving rise to superfluidity in neutron-star crusts is a highly non-local phenomenon involving both neutrons 
bound inside clusters and free neutrons. Using the band theory of solids, the critical temperature for superfluidity is found to be 
reduced by $\sim 10-20\%$ as compared to pure neutron matter. Non-local effects have a much stronger impact on superfluid dynamics. 

Due to non-dissipative entrainment effects, some regions of the crust are found to strongly resist the neutron superfluid flow. 
This may have implications for various astrophysical phenomena. For instance, 
the amount of angular momentum that the neutron superfluid can possibly transfer to the crust is limited by entrainment, thus challenging the interpretation of large pulsar 
glitches~\citep{cc06}. Because the neutron superfluid is coupled to the crust, entrainment leaves its imprint on the spectrum of 
collective excitations and their damping~\citep{pethick2010,cirigliano2011}. This could affect the thermal relaxation of the crust 
which has been recently monitored in several persistent soft X-ray transients. 

\acknowledgments This work was supported by FNRS (Belgium), NSERC (Canada) and the Research Networking Programme CompStar of the European 
Science Foundation.

\bibliography{chamel_talk}

\end{document}